# Drag-Controlled Regime Transitions in the Eddy Saturation Mechanism of the Antarctic Circumpolar Current

**Takuro Matsuta[1], Yuki Tanaka[2] and Atsushi Kubokawa[1]**

[1] Faculty of Environmental Earth Science, Hokkaido University, Sapporo, Japan.

[2] Department of Ocean Sciences, Tokyo University of Marine Science and Technology, Minato-ku, Japan

Corresponding author: Takuro Matsuta (matsuta@ees.hokudai.ac.jp)

**Key Points:**

- Eddy saturation in the Antarctic Circumpolar Current occurs across a wide range of bottom-drag regimes.
- The eddy saturation mechanism is regulated by strength of wind stress relative to drag coefficient.
- Drag-controlled regime transitions determine how the Antarctic Circumpolar Current responds to wind forcing.

## Abstract

Eddy saturation—the weak sensitivity of Antarctic Circumpolar Current (ACC) transport to wind stress—is a fundamental feature of Southern Ocean dynamics, yet the processes that maintain this state remain debated. Previous studies have proposed different mechanisms, including adjustments of eddy diffusivity and standing meanders, but the conditions under which each mechanism dominates are unclear. Here we use an idealized reentrant channel model to examine how drag strength controls the eddy saturation. When the wind strength relative to friction is below a certain threshold, eddy saturation is governed by a combination of standing meander and eddy diffusivity adjustments; once the threshold is exceeded, it is governed solely by standing meander adjustment. These results suggest that changes in drag strength may account for the divergent eddy saturation mechanisms reported across studies.

## Plain Language Summary

The Antarctic Circumpolar Current (ACC) is the world's strongest ocean current and plays an important role in Earth's climate. Surprisingly, many ocean models show that the ACC does not become much stronger even when winds increase. This behavior, called eddy saturation, occurs because ocean eddies adjust to absorb the extra wind energy. However, previous studies have proposed different explanations for how this adjustment happens. In this study, we found that eddy saturation occurs under a wide range of conditions, but the mechanism that maintains it changes depending on the strength of bottom friction. These results show that accurately representing seafloor friction is important for predicting how the Southern Ocean will respond to changing winds.

## 1 Introduction

The Antarctic Circumpolar Current (ACC) is a unique current that connects the three ocean basins without lateral obstacles in the latitude of the Drake Passage. This unblocked structure makes the ACC a critical conduit for global ocean tracer exchange (Talley, 2013). Since the mid-twentieth century, the Southern Hemisphere westerlies driving the ACC have strengthened and shifted poleward due to global warming (Fogt & Marshall, 2020). Understanding how the ACC transport responds to these changes is an important issue for diagnosing oceanic tracer circulation.

Despite increasing wind stress, observations suggest that the zonal ACC transport—characterized by its baroclinicity (meridional isopycnal slope)—has exhibited no significant trend since the mid-twentieth century (Böning et al., 2008). This *eddy saturation* phenomenon is supported by high-resolution models (e.g., Bishop et al., 2016; Constantinou & Hogg, 2019; Robert Hallberg & Gnanadesikan, 2006; Matsuta et al., 2024; David R. Munday et al., 2013; Spence et al., 2010). Traditionally, eddy saturation is explained by the balance between wind stress and eddy interfacial form stress (EIFS) associated with baroclinic instability (Stewart et al., 2023 and references therein). According to J. Marshall & Radko (2003), EIFS scales with $\kappa s$, where $\kappa$ is Gent–McWilliams (GM) eddy diffusivity and $s$ is the baroclinicity. If the GM eddy diffusivity increases proportionally with wind stress, the baroclinicity remains unchanged. This mechanism has been supported by experiments showing that baroclinicity is sensitive to wind in coarse-resolution models with constant $\kappa$, but less sensitive in coarse-resolution model with variable $\kappa$ (Gent & Danabasoglu, 2011; Mak et al., 2018).

However, recent studies suggest that standing meanders may be as vital as transient eddies (Katsumata, 2017; Stewart et al., 2023; Thompson & Naveira Garabato, 2014). Topographically forced standing meanders produce a net heat transport and thus, like baroclinic instability, contribute to the reduction of baroclinicity (R. Hallberg & Gnanadesikan, 2001). This implies that eddy saturation can be sustained through the standing meander adjustment rather than solely through changes in eddy diffusivity. Supporting this, some studies (Farneti et al., 2015; Kong & Jansen, 2021) showed that eddy saturation can emerge even in coarse-resolution models with constant $\kappa$ —a result that appears inconsistent with previous findings (Gent & Danabasoglu, 2011; Mak et al., 2018).

It remains unclear whether transient eddies or standing meanders play the dominant role in eddy saturation. Stewart et al. (2023) suggested that the choice of bottom drag parameterization may switch the underlying mechanism. Specifically, under a quadratic drag scheme, standing meander adjustment appears responsible for eddy saturation. In contrast, under a linear scheme, GM diffusivity adjustments dominate in weak-wind regimes, while standing meanders prevail under strong winds. Because eddy dynamics differ between linear and quadratic drag formulations (Arbic & Scott, 2008; Chen, 2023; Gallet & Ferrari, 2020), this hypothesis has some physical plausibility. However, because linear drag is typically more dissipative than quadratic drag (Arbic & Scott, 2008), it is uncertain whether these differences stem from the mathematical form of the drag or simply from differences in total frictional strength. To address this gap, we investigate the eddy saturation mechanism using an idealized reentrant channel model, systematically varying the strength of linear friction. Our results suggest that often-overlooked dissipation processes play a key role in controlling large-scale circulation and mesoscale dynamics, underscoring the importance of accurate dissipation parameterization.

## 2 Theoretical background

In the zonally-averaged framework, the momentum imposed by westerlies at the ocean surface is transferred downward by eddy interfacial form stress (EIFS), standing-meander interfacial form stress (SIFS), and Coriolis force associated with residual circulation:

$$\underbrace{\rho_0 f \frac{\left[\overline{v'\rho'}\right]}{\partial_z \rho_{bg}}}_{\text{EIFS}} + \underbrace{\rho_0 f \frac{\left[\overline{v}^{\dagger}\overline{\rho}^{\dagger}\right]}{\partial_z \rho_{bg}}}_{\text{SIFS}} - \underbrace{\rho_0 f \psi_{res}}_{\text{RES}} \approx \tau_w, \qquad (1)$$

where $\rho_0$ is the reference density, $f$ is the Coriolis parameter, $\rho_{bg}$ is the background density defined as horizontally- and temporally- averaged density, $\partial_{\#}$ indicated the partial derivative by subscript #, $v$ is the meridional velocity, $\rho$ is the potential density, $\psi_{res}$ is the residual circulation, and $\tau_w$ is the wind stress (J. Marshall & Radko, 2003). The temporally-averaged component of variables is defined as *mean,* $\bar{\cdot}$, and eddy component $\cdot'$ is defined as the deviation from the mean. Similarly, standing component $\cdot^{\dagger}$ is defined as the deviation from a zonal-averaged value $[\cdot]$. EIFS represents the interfacial form stress associated with baroclinic instability, whereas SIFS represents the interfacial form stress associated with standing meanders. Positive values correspond to a transfer of eastward momentum toward the ocean bottom. In the GM parameterization framework (Gent & McWilliams, 1990), the EIFS is parameterized by the GM eddy diffusivity as follows,

$$\frac{\left[\overline{v'\rho'}\right]}{\partial_z \rho_{bg}} \approx \kappa_{GM} s, \tag{2}$$

where $\kappa_{GM}$ is the GM eddy diffusivity constant, and $s = -\partial_y[\overline{\rho}]/\partial_z \rho_{bg}$ is the baroclinicity. In the eddy-saturated regime, $s$ is approximately constant; therefore, an increase in EIFS with wind forcing is equivalent to an increase in $\kappa_{GM}$ with wind. By comparing the responses of EIFS and SIFS to changes in the westerlies, we assess the relative contributions of eddy diffusivity and standing meander adjustments to eddy saturation.

## 3 Configuration of numerical experiments

The numerical model is an idealized eddy-permitting $\beta$-plane reentrant channel implemented by MITgcm (J. Marshall et al., 1997). Here, we briefly describe the model configuration (Figure 1). The detailed setup is described in Matsuta et al. (2026). The domain is 4000-km long (zonal, x-direction), 2000-km wide (meridional, y-direction), and 3881-m deep (vertical, z-direction). The Coriolis parameter is $f = f_0 + \beta y$, with $f_0 = -10^{-4}\ \mathrm{s}^{-1}$ and $\beta = 10^{-11}\ \mathrm{m}^{-1}\mathrm{s}^{-1}$. A single Gaussian ridge whose height is 1500 m and zonal scale is 300 km is located. We used a Cartesian grid with a horizontal resolution of $10 \times 10\ \mathrm{km}$. There are 29 vertical levels increasing in thickness from 8.5 m at the ocean surface to 248 m at depth. At the southern boundary, the free-slip condition is imposed and there is no heat transport. The northern boundary is also free slip, but a sponge layer is imposed within 100 km of the northern boundary, where temperature is relaxed toward an exponential profile ranging from 8°C at the surface to 0°C at the bottom with a scale height of 1000 m. The initial temperature is also given by this profile. The density is prescribed by a temperature-only linear equation of state with a thermal expansion coefficient of $2.0 \times 10^{-4}\ ^{\circ}\mathrm{C}^{-1}$. The subgrid horizontal eddy viscosity is $100\ \mathrm{m}^2\ \mathrm{s}^{-1}$ and horizontal eddy diffusivity is zero. The vertical eddy viscosity and diffusivity is weak in the ocean interior; the K-profile parameterization (KPP) mixing scheme (Large et al., 1994) is applied to represent a surface mixed layer.

We conduct a set of experiments with four different values of linear bottom drag ($r = \{10^{-5}, 5.0 \times 10^{-4}, 10^{-3}, 10^{-2}\}\ \mathrm{m\ s}^{-1}$), hereafter referred to as the LOW, MEDIUM−, MEDIUM+, and HIGH cases. Changing the linear drag can be interpreted as changing the effective roughness of the ocean bottom (Klymak et al., 2021). In each configuration, the strength of the westerlies is changed to examine the responses of interfacial form stress. The wind forcing is prescribed as follows.

$$\tau_w(y) = \tau_0 \sin\left(\pi \frac{y}{L_y}\right), \quad 0 \leq y \leq L_y, \tag{3}$$

where $\tau_0 = \{0.05, 0.1, 0.2, 0.3, 0.4\}\ \mathrm{N\ m}^{-2}$ is the wind strength, and $L_y = 2000\ \mathrm{km}$ is the meridional extent of the domain.

The model is integrated for 100 years with the final 30 years utilized for calculations. We define an averaged value of this period as *mean* and a deviation from mean as *eddy.* It should be noted that reaching a thermal equilibrium state would require integrations of roughly several hundred years under the high-drag cases (D. R. Munday et al., 2015), but this is not computationally feasible. Therefore, it is unclear whether the residual circulation term in $(1)$ results from heat input through the sponge layer or from the system not having reached thermal equilibrium, although the residual circulation term is consistently smaller than EIFS and SIFS in all cases except for the case of $\tau_0 = 0.05\ \mathrm{N\ m}^{-2}$. We therefore do not discuss the residual circulation term in this study.

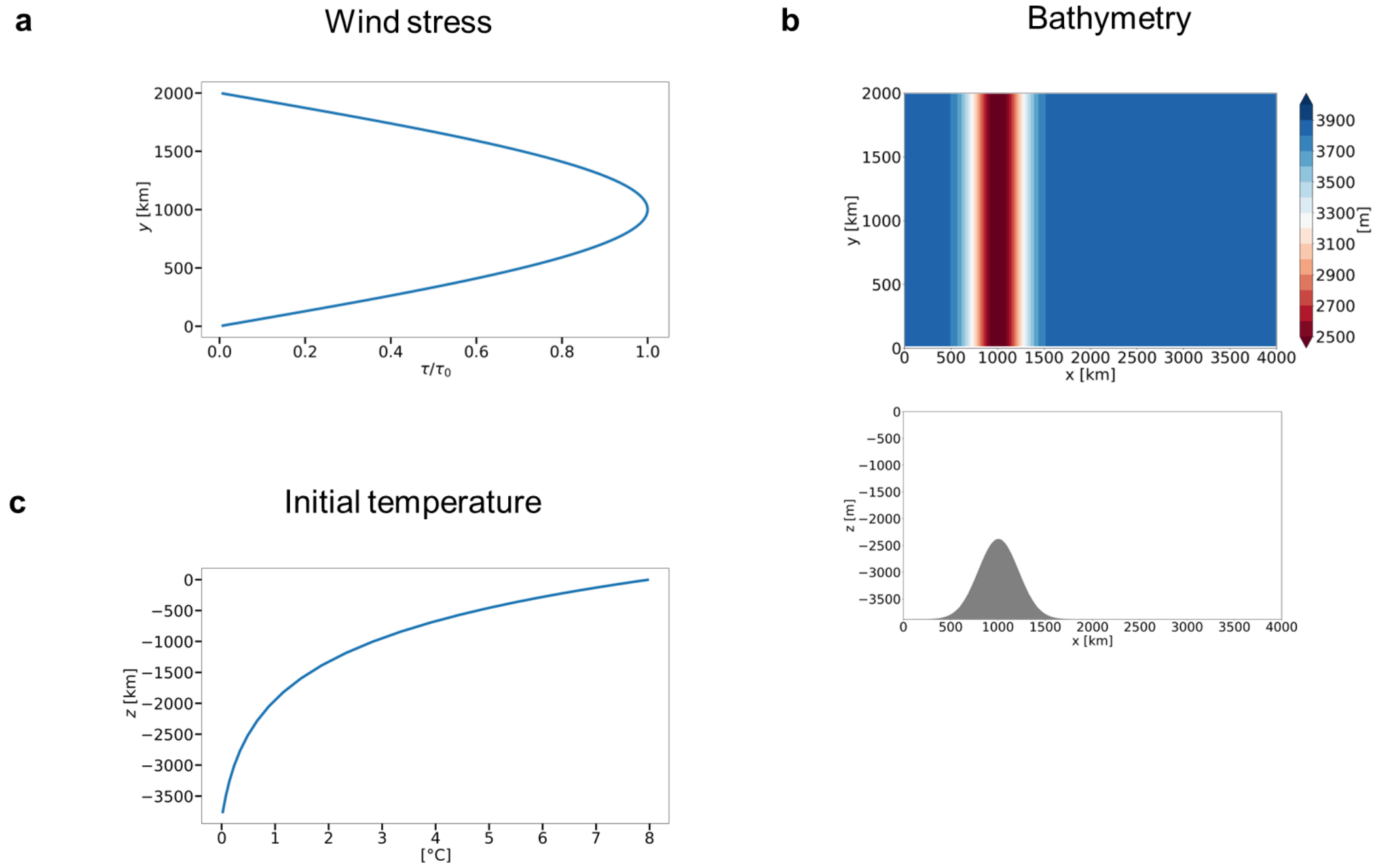

**Figure 1 Summary of model configuration. (a) The meridional structure of wind stress normalized by the maximum value. (b) Horizontal distribution (upper) and x-z section (lower) of bathymetry. (c) The vertical profile of initial temperature.**

## 4 Results

Figure 2 illustrates the horizontal distributions of the barotropic streamfunction and eddy kinetic energy (EKE) for representative cases. Here, the barotropic streamfunction is defined as

$$\psi(x,y) = \int_0^y \int_{-h}^0 \overline{u}\, dzdy\,, \tag{4}$$

and EKE is defined

$$\mathrm{EKE} = \frac{1}{2}\left(\overline{u'^2} + \overline{v'^2}\right), \tag{5}$$

where $u$ is the zonal velocity. Figure 2 shows that EKE increases with wind stress for each configuration. Previous studies (Mak et al., 2017; D. P. Marshall et al., 2017; Matsuta et al., 2026) suggest that the increase in EKE with stronger wind forcing reflects a tendency for enhanced eddy activity to dissipate the excess energy input by the wind, thereby maintaining an eddy-saturated state.

The barotropic streamfunctions reveal that standing meanders and gyre circulations weaken as the drag coefficient increases. Under weak wind stress of $\tau_0 = 0.1 \mathrm{~N~m^{-2}}$, gyre circulations is prominent in LOW, whereas they are not evident in the other cases. The spatial scale of standing meanders also scales inversely with friction. As wind stress intensifies, wind-driven gyres and standing meanders expand across all cases. This response is most pronounced in the MEDIUM− and MEDIUM+ cases, where distinct wind-driven double-gyre structures emerge only under strong wind forcing. This behavior is consistent with the response of the vertical structure. Figure 3a shows the fraction of barotropic mode in the mean kinetic energy averaged over the domain. The flow becomes increasingly barotropic with decreasing friction or increasing wind. As the barotropic mode becomes more dominant, the influence of the ridge increases, making it easier to satisfy Sverdrup balance (Jouanno & Capet, 2020; Nadeau & Ferrari, 2015).

Figure 3b displays the relationship between wind stress and baroclinicity, defined as the zonal-mean $1\,°\mathrm{C}$ isotherm depth difference between y=500 km and y=1500 km. While baroclinicity increases with the drag coefficient following the frictional control relation (Mak et al., 2017; D. P. Marshall et al., 2017; Matsuta et al., 2026), it remains remarkably insensitive to wind stress with $\tau_0 \geq 0.2 \mathrm{~N~m^{-2}}$ across all friction regimes. This result demonstrates that eddy saturation is achieved irrespective of the magnitude of friction.

To investigate the mechanisms sustaining eddy saturation, we decompose the interfacial form stress into EIFS and SIFS. Given those weak vertical variations, we evaluate the domain-averaged EIFS and SIFS values at 1400 m, well below the mixed layer and above the bottom topography. Although the domain averaged EIFS is contaminated by the eddy flux along the sponge layer at the northern boundary, we have confirmed that restricting the analysis to the interior region (from y = 250 km to y = 1750 km) does not affect the qualitative conclusions. This also applies to the analysis of baroclinic energy conversion presented in Section 5. Figure 3c reveals distinct EIFS responses across the drag regimes. Although EKE increases with increasing wind stress in the LOW case (Figure 2), EIFS remains negligible. In the HIGH case, EIFS increases with wind forcing. However, the increase in EIFS (and eddy diffusivity) is sublinear, indicating that eddy diffusivity does not solely explain the eddy saturation. The MEDIUM− and MEDIUM+ cases exhibits distinct behavior: EIFS increases with wind stress under the weak wind regime, but decreases under the strong wind regime. In contrast to EIFS, the response of SIFS is relatively simple: SIFS increases almost linearly with wind forcing in all cases (Figure 3d).

The response of EIFS can be characterized by plotting it against *relative wind strength* normalized by the drag coefficient and reference density, $\hat{\tau} := \tau_0/\rho_0 r$. As shown in Figure 4a, EIFS decreases with increasing wind in the regime where $\hat{\tau} > 2.0 \times 10^{-1}\ \mathrm{m\ s^{-1}}$ whereas EIFS increases with $\hat{\tau}$ below $\hat{\tau} = 2.0 \times 10^{-1}\ \mathrm{m\ s^{-1}}$. Conversely, the response of SIFS is not sensitive to $\hat{\tau}$ (Figure 4b).

These results suggest that standing meander adjustment contributes to eddy saturation across a wide range of wind and frictional conditions. When the wind is strong relative to friction, eddy diffusivity adjustment is negligible, and eddy saturation can be explained by standing meanders alone. In contrast, when the wind is weak relative to friction, the regulation of eddy diffusivity cannot be neglected, and eddy saturation is achieved through a combination of eddy and standing meander adjustments. In models with intermediate friction, increasing wind forcing causes the system to cross the threshold, shifting from eddy saturation governed by both eddies and standing meanders to a regime dominated solely by standing meander adjustment.

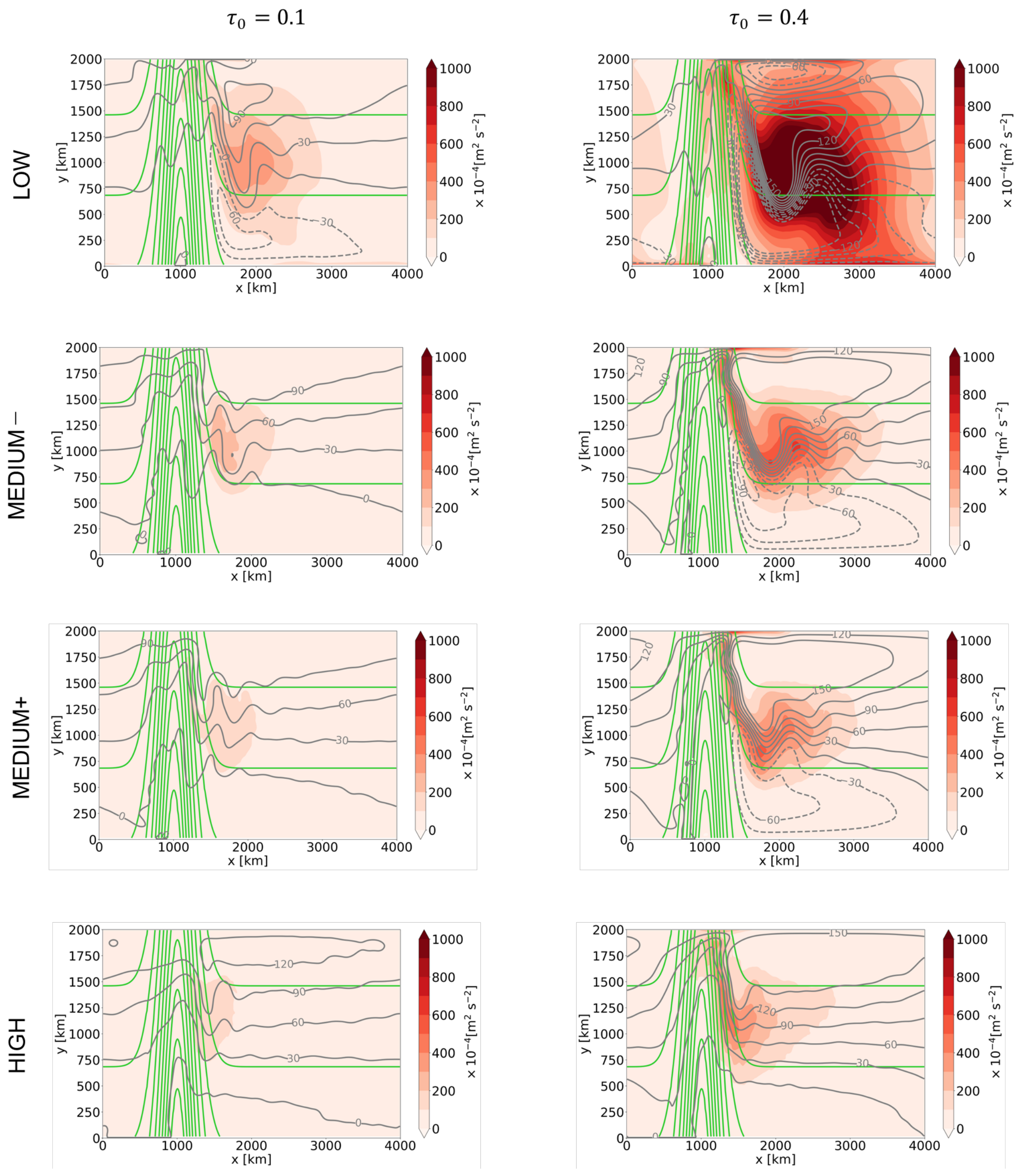

**Figure 2 Sensitivity of horizontal distribution of barotropic streamfunction (gray contour) and vertically-averaged EKE (color shade) on bottom drag and wind stress. From top to bottom, the panels show the LOW, MEDIUM−, MEDIUM+, and HIGH cases. The left column corresponds to the case of $\tau_0 = 0.1\ \mathrm{N\ m^{-2}}$, and the right column corresponds to the case $\tau_0 = 0.4\ \mathrm{N\ m^{-2}}$. The streamfunction contours whose magnitude are larger than 150 Sv are not shown. Green lines are geostrophic contours with a contour interval of $2.0 \times 10^{-9}\ \mathrm{m^{-1}\ s^{-1}}$. Here, the geostrophic contour is defined as $-f/h$, where $h$ is the ocean thickness.**

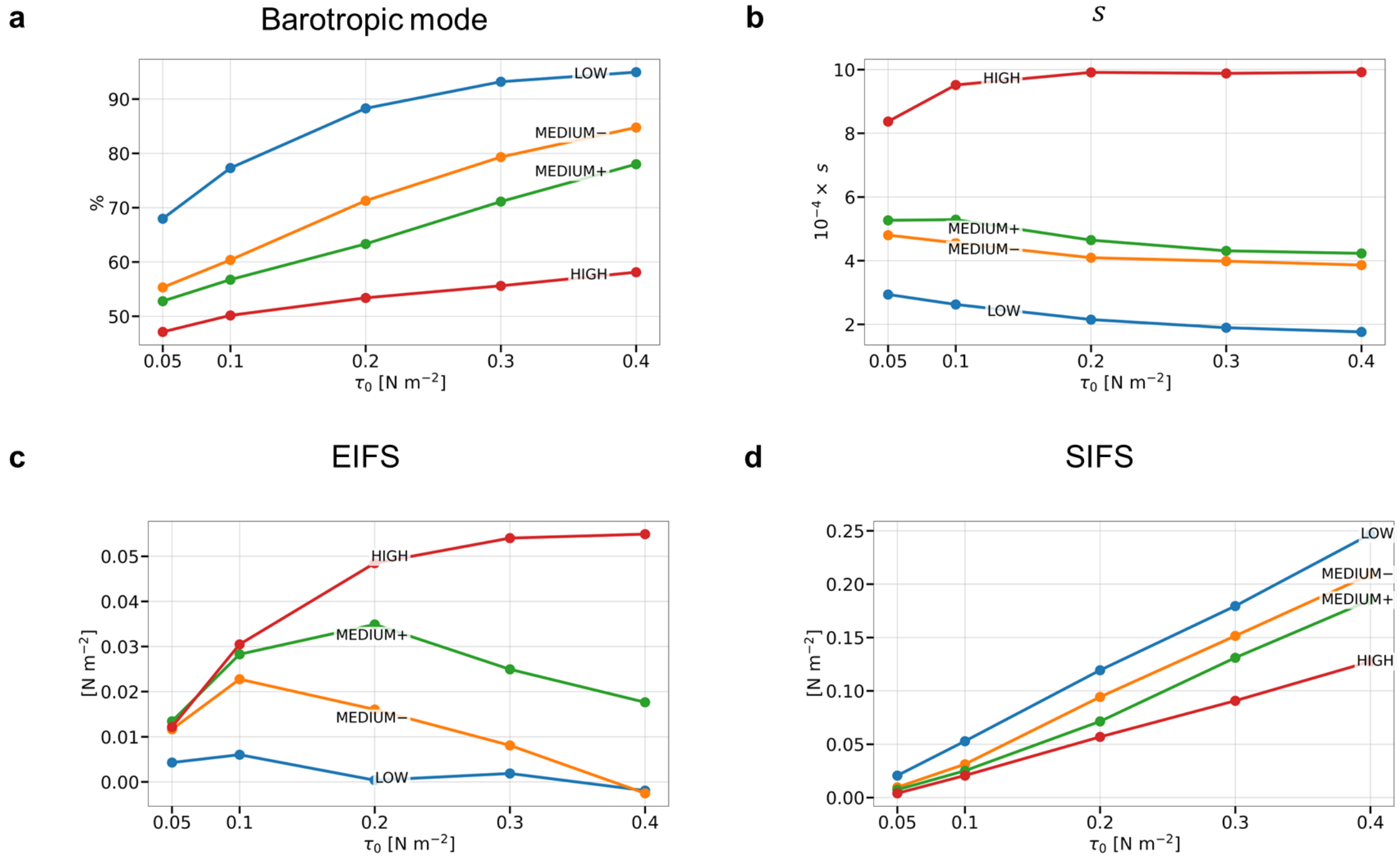

**Figure 3 Sensitivity of (a) barotropic mode fraction, (b) baroclinicity, (c) EIFS, and (d) SIFS to wind strength. Blue, orange, green and red curves are for LOW, MEDIUM−, MEDIUM+, and HIGH cases.**

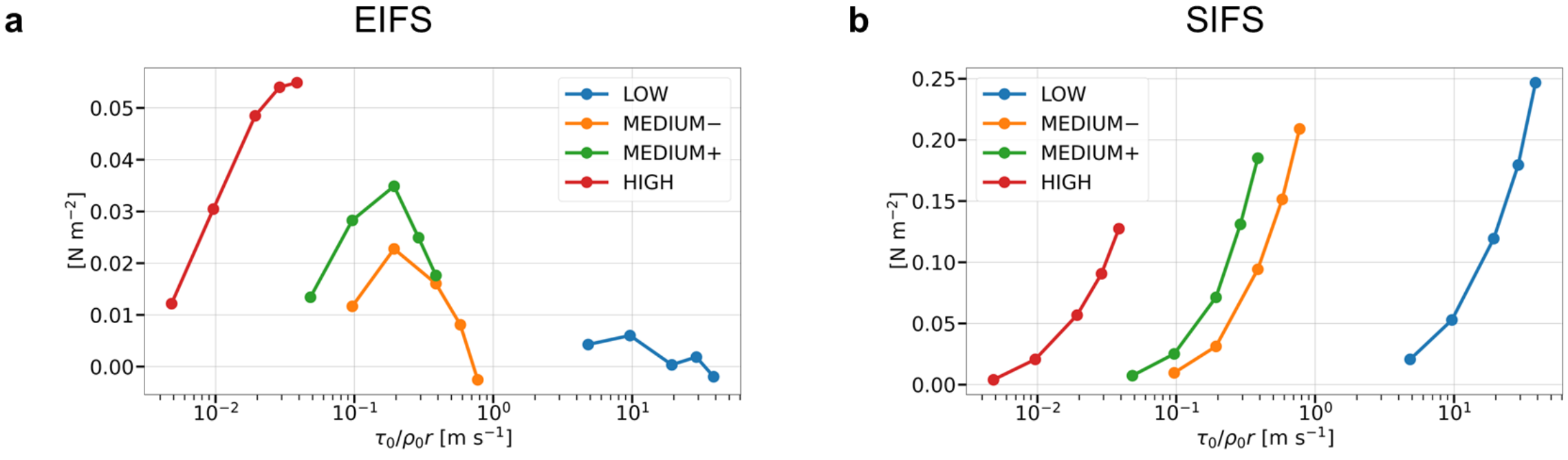

**Figure 4 Sensitivity of (a) EIFS and (b) SIFS to $\hat{\tau} = \tau_0/\rho_0 r$. Blue, orange, green and red curves are for LOW, MEDIUM−, MEDIUM+, and HIGH cases.**

## 5 A complementary view from eddy energetics

From a momentum perspective, standing meander adjustment is more important than baroclinic instability in eddy saturation; however, from an energy perspective, baroclinic instability plays a crucial role in maintaining the eddy saturated state. To demonstrate this, we analyze the energy conversion terms. The baroclinic energy conversion rate (BCR) and barotropic energy conversion rate (BTR) are given by the following equations,

$$\mathrm{BCR} = \frac{g}{\partial_z \rho_{bg}(z)} \overline{\rho' \mathbf{u}'_h} \cdot \nabla_h \overline{\rho} = \underbrace{\frac{g}{\partial_z \rho_{bg}(z)} \overline{\rho' u'} \frac{\partial \overline{\rho}}{\partial x}}_{\mathrm{BCR}_x} + \underbrace{\frac{g}{\partial_z \rho_{bg}(z)} \overline{\rho' v'} \frac{\partial \overline{\rho}}{\partial y}}_{\mathrm{BCR}_y}, \tag{6}$$

and

$$\mathrm{BTR} = -\rho_0 \left( \overline{u' \mathbf{u}'} \cdot \nabla \overline{u} + \overline{v' \mathbf{u}'} \cdot \nabla \overline{v} \right), \tag{7}$$

where $\mathbf{u} = (u, v, w)$ is the three-dimensional velocity, $\nabla$ is the three-dimensional operator, and $\nabla_h$ is the horizontal component of $\nabla$ (Olbers et al., 2012). Positive BCR and BTR correspond to EKE generation through baroclinic and barotropic instability, respectively. According to Figure 5a, the domain-integrated BCR increases with $\hat{\tau}$ in all cases. This result suggests that, even if EIFS becomes less important than SIFS in the vertical momentum transfer, baroclinic instability plays roles. To examine changes in the nature of BCR, Figure 5b shows the fractional contribution of $\mathrm{BCR}_x$, i.e., the zonal component of BCR to BCR. In the HIGH case, the contribution of $\mathrm{BCR}_x$ remains around 30% even as the wind strengthens, indicating that the meridional eddy flux is the primary driver of baroclinic instability. Indeed, the vertically-integrated BCR in HIGH with $\tau_0 = 0.4\ \mathrm{N\ m^{-2}}$ (Figure 5c) takes large positive values not only along the standing meander but also along the zonal jet away from topography. In contrast, in the regime beyond the threshold of $\hat{\tau} = 2.0 \times 10^{-1}\ \mathrm{m\ s^{-1}}$ (i.e., in the regime where EIFS is negligible), the contribution of $\mathrm{BCR}_x$ increases with strengthening westerlies (although it slightly decreases at τ = 0.4 in the MEDIUM− case). Consistently, as shown in Figure 5d, BCR of MEDIUM+ with $\tau_0 = 0.4\ \mathrm{N\ m^{-2}}$ takes large values only within the southward flow downstream of the topography, indicating that the zonal component of eddy flux becomes important for BCR. This baroclinic instability transfers kinetic energy to the bottom layer, enabling the southward flow to feel the bottom topography and thereby allowing SIFS to become effective. It is also noteworthy that negative BCR appears in the downstream region (east of x = 2000 km). In this region, eddies act to enhance baroclinicity, and the presence of this upgradient eddy heat flux leads to very small EIFS for $\hat{\tau} > 2.0 \times 10^{-1}\ \mathrm{m\ s^{-1}}$.

It should be also noted that the response of BTR strongly depends on $\hat{\tau}$ (Figure 5e). When $\hat{\tau}$ is less than $2.0 \times 10^{-1}\ \mathrm{m\ s^{-1}}$, BTR shows little increase with wind forcing. In contrast, once this threshold is exceeded, BTR increases with wind. As a result, in this regime, EKE is not solely associated with baroclinic instability but is also generated by barotropic instability. Matsuta et al. (2024) pointed out that the sensitivity of BTR to wind forcing differs among models (c.f., Matsuta & Mitsudera, 2024; Wu et al., 2017; Youngs et al., 2017); our results suggest that these differences may be attributable to differences in frictional strength.

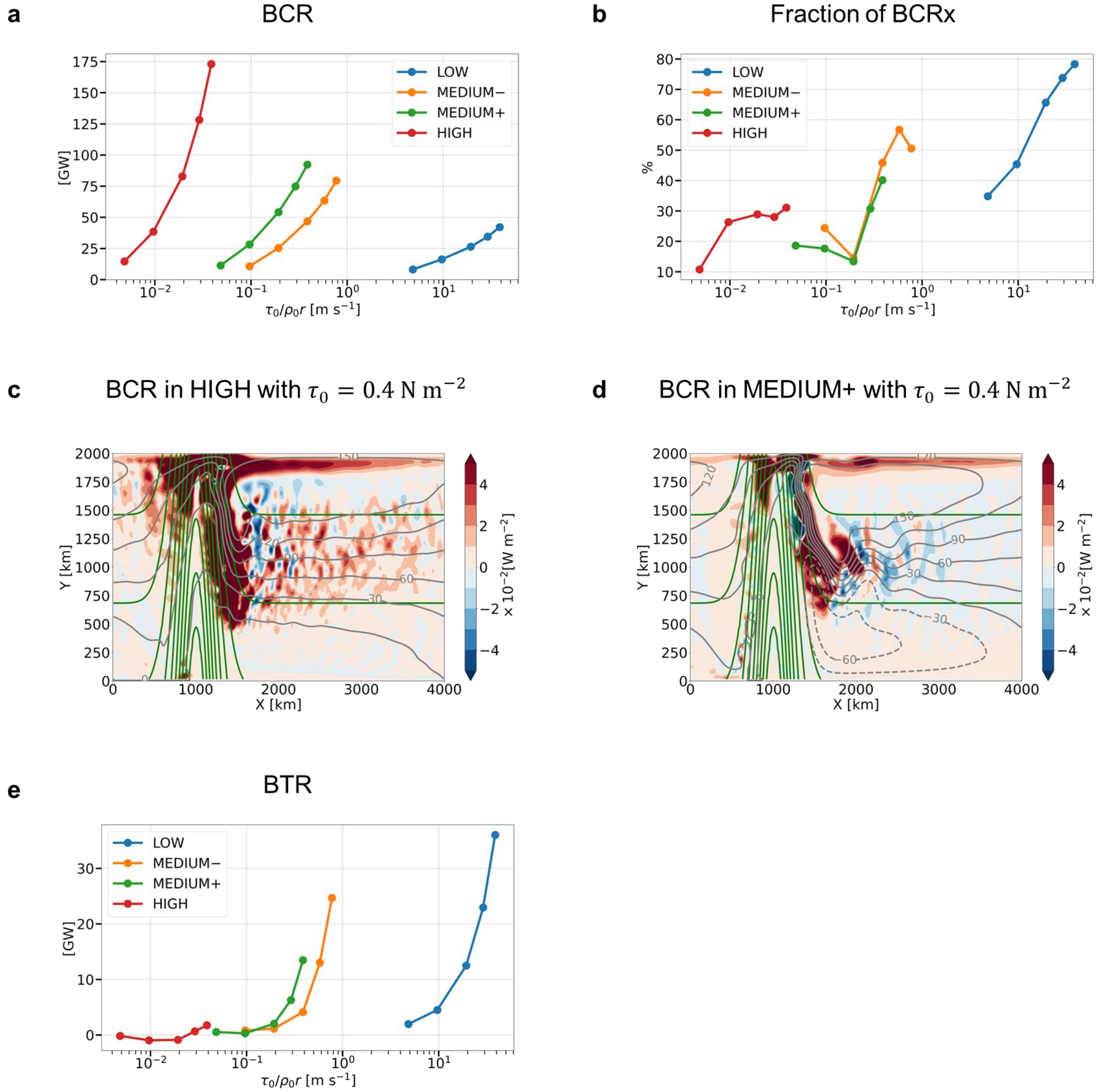

**Figure 5 Sensitivity of eddy energy conversion rates. (a) Sensitivity of domain-integrated BCR to $\hat{\tau} = \tau_0/\rho_0 r$. Blue, orange, green, and red curves are for LOW, MEDIUM−, MEDIUM+, and HIGH cases. (b) Same as (a) but for fraction of $\mathrm{BCR}_x$ to BCR. (c) Horizontal distribution of vertically-integrated BCR in HIGH with $\tau_0 = 0.4\ \mathrm{N\ m^{-2}}$. Green and gray contours are the same as those in Figure 2. (d) Same as (c) but in MEDIUM+ with $\tau_0 = 0.4\ \mathrm{N\ m^{-2}}$. (e) Same as (a) but for BTR.**

## 6 Conclusions and Discussions

We investigated the mechanisms of eddy saturation using an idealized reentrant channel model with changing linear bottom drag. While eddy saturation is achieved across all drag regimes, the underlying adjustment is fundamentally drag-dependent. In the regime where the wind strength relative to linear drag coefficient is below a threshold, both eddy diffusivity and standing meander adjustments are responsible for eddy saturation. Once the threshold is exceeded, eddy saturation can be explained solely by standing meander adjustment. In this regime, the interfacial form stress associated with baroclinic instability is negligible, and momentum transfer to the lower layer is dominated by standing Rossby waves. However, this result does not imply that baroclinic instability plays no role in this regime: baroclinic instability acts indirectly by mediating standing meander adjustment through the barotropization of mean flow. Consistent with this view, baroclinic energy conversion increases with wind regardless of bottom drag coefficient.

These findings provide a unified framework to reconcile disparate results in the literature. Studies emphasizing standing meanders (Kong & Jansen, 2021; Stewart et al., 2023) generally utilized weak-drag configurations, while those highlighting eddy diffusivity regulation employed stronger dissipation (Mak et al., 2018). Consequently, we suggest that changes in bottom drag strength, rather than the mathematical form of the drag, may dictate the specific mechanical pathway of eddy saturation.

Our results highlight that accurately estimating bottom drag is essential for reliably simulating future changes in the Southern Ocean circulation. Although the reduced sensitivity in circumpolar transport has been reported even in coarse simulations with constant GM parameter (Farneti et al., 2015; Kong & Jansen, 2021), it remains unclear whether this response represents eddy saturation in the real ocean. In the real ACC, strong friction associated with topographic roughness is present (Klymak et al., 2021; Mak et al., 2022); however, low-resolution models do not adequately represent this effect and may therefore overestimate the standing meander adjustments. In addition, wind changes themselves may alter the drag coefficient by exciting lee waves (Yang et al., 2023). In such models, the response of baroclinic instability exerts a strong influence on baroclinicity. These uncertainties underscore the importance of better constraints on bottom friction and the continued development of mesoscale eddy parameterizations with a variable GM coefficient to improve projections of the Southern Ocean response to changing winds.

## Acknowledgments

Takuro Matsuta was supported by JSPS KAKENHI Grant Number 24K17120. This work was supported in part by the Collaborative Research Program (2025S2-CD-1 and 2026S2-CD-2) of Research Institute for Applied Mechanics, Kyushu University, and the Cooperative Research Activities of Collaborative Use of Computing Facility Program (JURCAOSCFG25-06) of Atmosphere and Ocean Research Institute, The University of Tokyo. This research was conducted using the FUJITSU Supercomputer PRIMEHPC FX1000 and FUJITSU Server PRIMERGY GX2570 (Wisteria/BDEC-01) at the Information Technology Center, The University of Tokyo. The authors used ChatGPT (OpenAI) and Gemini (Google) for English language editing and improving clarity. All outputs were reviewed and edited by the authors, who take full responsibility for the content.

## Open Research

Model configuration, output and analysis scripts are available at https://doi.org/10.5281/zenodo.19969114. The MITgcm model is publicly available at https://mitgcm.readthedocs.io/en/latest/index.html.

## Conflict of Interest Disclosure

The authors declare there are no conflicts of interest for this manuscript.